\documentclass[12pt]{article}
\usepackage{epsfig,epsf}
\begin{document}

%\begin{center}
%{\Large {\bf The Discrete Asymptotically-Free BFKL Pomeron Fits HERA Data}}
%\end{center}
\begin{flushright}
CERN-PH-TH/2008-030\\
DESY 08-028\\
SHEP-08-08\\
%\date{\today}
\end{flushright}

\begin{center}
{\Large{\bf Evidence for the Discrete Asymptotically-Free\\ BFKL Pomeron from HERA Data}} \\
[0.5cm]

{\large J. Ellis~$^1$, H. Kowalski~$^2$ and D.A. Ross~$^{1,3}$} \\ [0.3cm]

{\it $^1$ Theory Division, Physics Department, CERN, CH-1211 Geneva 23, Switzerland}
\\[0.1cm]

{\it $^2$ Deutsches Elektronen-Synchrotron DESY, D-22607 Hamburg, Germany}\\[0.1cm]

{\it $^3$ School of Physics and Astronomy, University of Southampton}\\
{\it Highfield, Southampton SO17 1BJ, UK} \end{center}
\vspace*{3 cm}

\begin{center}
{\bf Abstract}
\end{center}

We show that the next-to-leading-order renormalization-group-improved
asymptotically-free BFKL Pomeron provides a good fit to HERA data on virtual
photoproduction at small $x$ and large $Q^2$. The leading discrete Pomeron
pole reproduces qualitatively the $Q^2$ dependence of the HERA data for $x \sim 10^{-3}$,
and a fit using the three leading discrete singularities reproduces
quantitatively the $Q^2$ and $x$ dependence of the HERA data for $x < 10^{-2}$.
This fit fixes the phase for all the BFKL wavefunctions at a chosen infrared scale.

\vspace*{3 cm}

\begin{flushleft}
CERN-PH-TH/2008-030\\
February 2008\\
\end{flushleft}

\newpage

%\section{Introduction}

The nature of the QCD Pomeron continues to perplex and intrigue both experimentalists
and theorists~\cite{FR}. Study of the Pomeron has been one of the most interesting
aspects of the HERA experimental programme, with the discovery of
a `hard' Pomeron in virtual photoproduction~\cite{hardP} whose relation to the `soft' Pomeron
that is familiar from traditional hadronic reactions~\cite{softP} is still the subject of
theoretical speculation. In the near future, the LHC will
provide  possibilities to test theoretical approaches that have been honed
with HERA data and may provide novel opportunities to study
new Pomeron physics.

Most of the HERA data on deep-inelastic structure functions are
described well by the asymptotically-free renormalization-group evolution
expressed in the Dokshitzer-Gribov-Lipatov-Altarelli-Parisi (DGLAP) \cite{DGLAP}
equations. On the other hand, it has been suggested that a more
appropriate framework for describing data at very low $x$ is the
diffusion in transverse momentum incarnated in the Balitsky-Fadin-Kuraev-Lipatov
(BFKL) equation~\cite{BFKL}.
There has been considerable discussion of HERA data at low $x$
in the context of unified BFKL and DGLAP  equations~\cite{cut}.
However, it has not yet been established whether  the {\it pure} BFKL Pomeron provides an
accurate description of data in the kinematic range accessible to HERA.

The BFKL equation with fixed strong coupling yields a leading Pomeron singularity
that is a Regge cut, not a pole.
Fixing the QCD coupling may be plausible in a suitable infra-red
limit, but the coupling runs significantly in the ranges of virtuality and transverse
parton momenta explored in inclusive measurements of structure functions
at HERA. Over 20 years ago, it was shown~\cite{lipatov86} that
within the BFKL formalism, the leading Pomeron
singularity
is  a discrete Regge pole if the strong QCD coupling is
treated correctly as asymptotically free and the infrared behaviour is encoded in a fixed phase.
This leading singularity is followed
by an infinite series of lower-lying poles.

The next-to-leading-order (NLO) QCD correction to the leading-order (LO)
asymptotical- ly-free BFKL equation is known~\cite{FL}, and it has been shown how
to re-sum higher-order corrections so as to tame the NLO corrections~\cite{salam}.
The asymptotically-free BFKL equation described in \cite{lipatov86}
describes {\it both} the $x$- dependence of the  un-integrated gluon distribution
and the dependence on the transverse momentum, $k$, of the gluon, and hence also the $Q^2$
dependence of the structure function at low $x$.
The only unknown quantity is the proton impact factor, $\Phi_p(k)$,
which describes the couplings of the proton to the Pomeron trajectories.
One benefit of the discrete Pomeron approach is that a simple expression for the un-integrated
gluon density can be obtained, in terms of a small number of parameters, by
expanding the proton impact factor in terms of the discrete set of solutions of the
asymptotically-free BFKL equation.
We recall that the BFKL Pomeron may be expressed as
an integral that includes DGLAP as a saddle-point approximation valid
in the double limit ${\rm ln}(1/x) \gg 1$ and $\alpha_s(Q^2) {\rm ln}(1/x) \ll 1$~\cite{saddle}.
However, this DGLAP approximation to the BFKL integral is no longer valid
when the second condition is not satisfied,  as in the case of  low-$x$ HERA data,
where the discrete series
of BFKL Pomeron Regge poles is  a better systematic approximation scheme.

We are unaware of any overall fit to HERA data made using the re-summed NLO
asymptotically-free BFKL Pomeron. We perform such a fit in this paper,
and show that it describes the inclusive virtual photoproduction  HERA
   data both
qualitatively and quantitatively. We show first that the leading BFKL Pomeron pole
provides a successful qualitative description of HERA data on inclusive virtual
photoproduction at small $x \sim 10^{-3}$, over a large range of $Q^2$. This fit
improves if lower-lying BFKL Pomeron poles are included, and we
show that the asymptotically-free BFKL approach provides an excellent quantitative fit to all
the inclusive HERA data at $x \le10^{-2}$, if the three leading BFKL Pomeron
Regge poles are included.
As well as the residues of the three BFKL Pomeron poles at zero momentum
transfer, $t$, the BFKL fit has an additional
free parameter corresponding to the value of the phase of the BFKL wave function
that is assumed to be fixed by infra-red dynamics at a momentum $k_0 \sim
0.3$~GeV~\footnote{The precise value of $k_0$ is not an essential parameter.}.

We consider first the BFKL analysis
of a zero-momentum-transfer process,
at fixed strong coupling, $\alpha_s$.
In this case, the eigenfunctions of the BFKL kernel are representations
of the two-dimensional conformal group in the space of the transverse
coordinates of the gluons, ${\rho}$. We include
the BFKL characteristic function up to NLO~\cite{FL}, and use the
re-summation of Scheme 3   proposed by Salam~\cite{salam}, which moderates
the correction to the leading intercept as well as preserving the sign of the
curvature of the characteristic function near the intercept, up to large
values of $\alpha_s$.

Considering only the leading conformal spin, the eigenfunctions may be written
in momentum space as
 \begin{equation} f_{\omega}(k^2) \ = \ \frac{\overline{f_\omega}(k)}{\sqrt{k^2}},
\end{equation}
with
 \begin{equation} \overline{f_\omega}(k) \ = \left(k^2\right)^{i\nu},
 \end{equation}
where the eigenvalue $\omega$ is the solution to the equation
 \begin{equation}
 \omega \ \equiv \ \chi(\alpha_s,\nu) \ = \
 \bar{\alpha}_s \left(1-A\bar{\alpha}_s \right)
\overline{\chi_0}\left(\frac{1}{2}+\bar{\alpha}_s B+\frac{\omega}{2}+i\nu\right)
 + \bar{\alpha}_s^2 \overline{\chi_1}(\nu). \label{findomega} \end{equation}

Here
$$ \bar{\alpha}_s \ \equiv \ \frac{C_A}{\pi} \alpha_s, $$
$$ \overline{\chi_0}(z)=2\left(\psi(1)-\Re e \left[\psi(z)\right] \right)$$
$$ A \ \equiv \ \frac{n_f}{36 C_A^3} \left(10 C_A^2+13 \right) -\frac{\pi^2}{6}, $$
and
$$ B \ = \ \frac{11}{8}-\frac{n_f}{12C_A^3} \left(C_A^2-2\right), $$
where $C_A = 3$ and $n_f$ is the number of active flavours at
momentum ${k}$. 
$\overline{\chi_1}(\nu)$ is the NLO characteristic function
given in~\cite{FL}, omitting the conformal symmetry-violating part
associated with the running of the coupling (which is subtracted so that
the ${\cal O}( \bar{\alpha}_s^2)$ terms on the RHS of
(\ref{findomega}) are not double-counted: see~\cite{salam}).
The implicit equation (\ref{findomega}) for $\omega$ is readily
solved using an appropriate combination of Newton's method and
iteration.

Turning now to the case of running coupling, it was shown in~\cite{lipatov86} that
the frequency $\nu$ of the oscillations
acquires a dependence
on $k$, such that for a fixed eigenvalue $\omega$,
$\nu_{\omega}(k)=\chi^{-1}(\omega,\alpha_s(k))$ is the solution to
\begin{equation} \omega \ = \chi\left(\alpha_s(k),\nu_\omega(k)\right).
\label{chiminus1}  \end{equation}
This leads immediately to a critical value of the transverse momentum,
$k_{\mathrm{crit}}$, such that
\begin{equation}
\omega \ = \ \chi(\alpha_s(k_{\mathrm{crit}}), 0). \label{kcrit}
\end{equation}
Provided $\chi^{\prime\prime}(\alpha_s(k_{\mathrm{crit}}), 0)$
is negative \footnote{ We use the notations
$\chi^{\prime\prime}(\alpha_s,\nu) \equiv
d^2\chi(\alpha_s,\nu)/d\nu^2$ and $\dot{\chi}(\alpha_s,\nu) \equiv
d\chi(\alpha_s,\nu)/d\alpha_s$.}, the value of $\nu_\omega(k)$ becomes imaginary  for
$k \, > \, k_{\mathrm{crit}}$ and the eigenfunction decreases
exponentially as $k \, \to \, \infty$. It is in order to ensure that
 $\chi^{\prime\prime}(\alpha_s(k_{\mathrm{crit}}), 0)$ remains negative
that we have opted for Scheme 3 of the re-summation procedure described
in~\cite{salam}.

For  $k \, \sim \, k_{\mathrm{crit}}$,
 the BFKL equation may be approximated as
\begin{equation}
 \left[ \frac{d^2}{d[\ln(k^2/k_{\mathrm{crit}}^2)]^2}+\frac{\beta_0}{2\pi}
 \frac{\dot{\chi}(\alpha_s(k_{\mathrm{crit}}),0)}
{\chi^{\prime\prime}(\alpha_s(k_{\mathrm{crit}}),0)}
 \ln\left(\frac{k^2}{k_{\mathrm{crit}}^2}\right)
 \right]
\overline{f_\omega}(k) \ = \ 0, \end{equation}
with
$$ \beta_0 \ = \frac{11C_A}{3}-\frac{2}{3}n_f .$$
We recognize this as Airy's equation with argument
proportional to $\ln(k^2/k_{\mathrm{crit}}^2)$.
Away from $k_{\mathrm{crit}}$, provided the running of the coupling
is not too fast, so that
$$ \frac{d \nu_\omega(k)}{d\ln(k^2)} \ \ll \ \nu_\omega(k), $$
the BFKL equation may be approximated semi-classically by
\begin{equation}
\left[ i \frac{d}{d\ln(k^2)}+\nu_\omega(k) \right]
\overline{f_\omega}(k) \ = \ 0, \label{semiclassical}\end{equation}
which has the solutions
\begin{equation}
\overline{f_\omega}(k) \ = \ e^{\pm i \varphi_\omega(k)} ,
\end{equation}
where
\begin{equation}
 \varphi_\omega(k) \ = \ 2\int_k^{k_{\mathrm{crit}}}
 \frac{dk^{\prime}}{k^{\prime} }
\left|  \nu_\omega(k)\right| .
\end{equation}

In all regions, the solutions {\it decrease} as $k \, \to \, \infty$,
and are well approximated   by
\begin{equation}
\overline{f_\omega}(k) =  \sqrt{3}  \sqrt[3]{\varphi_\omega(k)}  K_{\frac{1}{3}}(\varphi_\omega(k))
\ \ \  \ \ (k \, > \, k_{\mathrm{crit}}) ,
\end{equation}
whereas
\begin{equation}
\overline{f_\omega}(k) = \sqrt[3]{\varphi_\omega(k)}
 \left[ J_{\frac{1}{3}}(\varphi_\omega(k))
+J_{-\frac{1}{3}}(\varphi_\omega(k))\right] \ \ \  \ \ (k \, < \, k_{\mathrm{crit}}),
\end{equation}
where we have expressed the appropriate Airy function in terms of
the modified Bessel function of the second kind, $K_{\frac{1}{3}}$,
and Bessel functions of the first kind, $J_{\pm\frac{1}{3}}$.
Away from $k \, \sim \, k_{crit}$ where $\varphi_\omega$ becomes large,
these Bessel function solutions approximate the solution to the 
semi-classical equation (\ref{semiclassical}).

It is important to note that the matching of the solutions at $k=k_{\mathrm{crit}}$
determines the phase of the oscillations in the region where $k < k_{\mathrm{crit}}$,
for a given value of $\omega$.
Following~\cite{lipatov86}, we encode the unknown infrared
behaviour of QCD  by assuming that it leads to a
fixed phase, $\eta$, at some low value of the transverse momentum,
$k_0$, which we take for definiteness to be 0.3 GeV~\footnote{ A change in this
 value of the infrared momentum scale can be compensated by a change
in the phase, $\eta$, so that the infrared
behaviour of QCD is in fact encoded using a single parameter.}. More precisely, the
infrared condition is given by
\begin{equation} \varphi_\omega(k_0) \ \equiv \
  2\int_{k_0}^{k_{\mathrm{crit}}}
 \frac{dk^{\prime}}{k^{\prime} }
\left|  \nu_\omega(k)\right|  \ = \
\left(n-\frac{1}{4}\right)\pi+\eta, \label{phasecond} \end{equation}
and means that, just above $k=k_0$, the wavefunction behaves like
\begin{equation}
\overline{f_\omega}(k) \ \sim \
\sin\left(\frac{\nu_\omega(k_0)}{k_0^2} \left(k^2-k_0^2\right)-\eta \right).
\label{phasecond2}
\end{equation}
Once the phase condition (\ref{phasecond}) is imposed, only a discrete set of values of
the eigenvalue $\omega$ are allowed simultaneously by the infrared phase condition
{\it and}  the phase condition imposed by the matching, giving rise to a description of the
QCD Pomeron as a discrete set of isolated poles, as opposed to the cut found if the
running of the strong coupling is neglected.

In order to express the low-$x$ structure function of the proton, $F_2(x,Q^2)$,
in terms of these eigenfunctions, the eigenfunctions themselves must be
convoluted with the impact factor $\Phi_p(k)$, that describes how the
proton couples to these trajectories at zero momentum transfer.
In the case of the un-integrated gluon density $x g(x,k)$, we have
\begin{equation}
x g(x,k) \ = \ \sum_n  \int \frac{dk^\prime}{k^\prime} \Phi_p(k^\prime)
 x^{-\omega_n} k^2  f^*_{\omega_n}(k^\prime) f_{\omega_n}(k),
\end{equation}
and the un-integrated gluon density is related to the structure function
by
\begin{equation}
F_2(x,Q^2) \ = \ \int_0^Q \frac{dk}{k} \Phi_{\mathrm{DIS}}(Q,k) x g(x,k),
\end{equation}
where the impact factor, $\Phi_{\mathrm{DIS}}$, that describes
 the coupling of the virtual photon to the trajectories is given by
(see \cite{FR})
\begin{equation}
 \Phi_{\mathrm{DIS}}(Q,k) \ = \ Q^2 \alpha_s(Q^2)
\sum_{q=1}^{n_f} e_q^2 \int_0^1 d\rho d\tau
\frac{1 - 2\rho(1-\rho) - 2\tau(1-\tau) + 12\rho(1-\rho)\tau(1-\tau)}
{Q^2\rho(1-\rho)+k^2\tau(1-\tau)}.
\end{equation}
The proton impact factor, $\Phi_p(k)$ is unknown {\it a priori} and has to be fit
to data. Since the eigenfunctions $f_{\omega_n}(k)$
 form an orthonormal set, we can expand the impact factor as a series in
these eigenfunctions with a discrete set of coefficients, $a_n$:
 \begin{equation} \Phi_p(k) \ = \sum_n a_n k^2  f_{\omega_n}(k),
\end{equation}
and exploit the orthogonality properties to write
\begin{equation}
 x g(x,k) \ = \  \sum_n  a_n x^{-\omega_n} k^2 f_{\omega_n}(k).
 \label{xgx}\end{equation}
A model for $\Phi_p(k)$ could be used to estimate the coefficients $a_n$, which could be also
constrained using other HERA data, e.g., on the diffractive production of vector mesons.

At sufficiently small $x$, we expect this sum to be dominated by the first few
poles. The contribution from the remaining  poles could be  approximated by assuming that
the effect of fixing the phase at $k_0$ on the allowed values of
$\omega$ is negligible for $\omega < 0.1$, and that the discrete set of eigenfunctions may be
replaced by a continuum. In this case, one simply adds to the expression
(\ref{xgx}) for the un-integrated gluon density the following integral that represents
the contribution from such a  continuum:
\begin{equation}
xg(x,k)^{\mathrm{(continuum)}} \ = \
 k \int_0^\infty d\nu \,
 b_p(\nu) \sin\left( \nu \ln\left(\frac{k^2}{k_0^2}\right)-\eta\right)
x^{-\chi(\alpha_s(k^2),\nu)} \theta\left(0.1-\chi(\alpha_s(k^2),\nu)\right),
 \label{continuum}
 \end{equation}
where $b_p(\nu)$ is a function that encodes the coupling of the proton to
all the remaining eigenfunctions, and is chosen to be real so that the
wavefunctions in this continuum also
respect the imposed infrared condition (\ref{phasecond2}).
In order to implement such a programme one would need additional parameters to characterize the
arbitrary function $b_p(\nu)$. However, in the region of $x$ and  $Q^2$ considered,
we find an excellent fit without making use
of such a continuum and the associated extra parameters, and hence do not consider it
further. Nevertheless, it
should be emphasized that at $Q^2$ that is sufficiently large for the  DGLAP analysis to
become valid, the double-leading-logarithm DGLAP behaviour would be embedded
mainly within this continuum contribution.

We have determined numerically the eigenfunctions of the leading four
poles of the NLO asymptotically-free BFKL Pomeron.
We limited ourselves to the first four poles because their  $\omega$ values are in the same range as the observed rate of rise,
 $\lambda$, of the $F_2$ measurements. This is determined by fitting the measured $F_2$ to $~x^{-\lambda}$ at fixed $Q^2$ and is
 closely related to the logarithmic derivative $d\log(F_2)/d\log(1/x)$. The values of $\lambda$ determined phenomenologically by
 experiment vary between $\lambda \approx 0.1$ for $Q^2 \le 0.6$ GeV$^2$ and $\lambda \approx 0.33$ for $Q^2 \ge 60$ GeV$^2$ ~\cite{ZH}.
 The leading eigenvalue, $\omega_1$, depends on the infrared phase, $\eta$, varying between $\omega_1=0.235$ at $\eta=0$
 and $\omega_1 = 0.315$ at $\eta=\pi/2$, The sub-leading eigenvalues are smaller, the fourth one, $\omega_4$, being $\approx 0.10$.

\begin{figure}
\vspace*{-3.0cm}
\centering
\epsfig{file=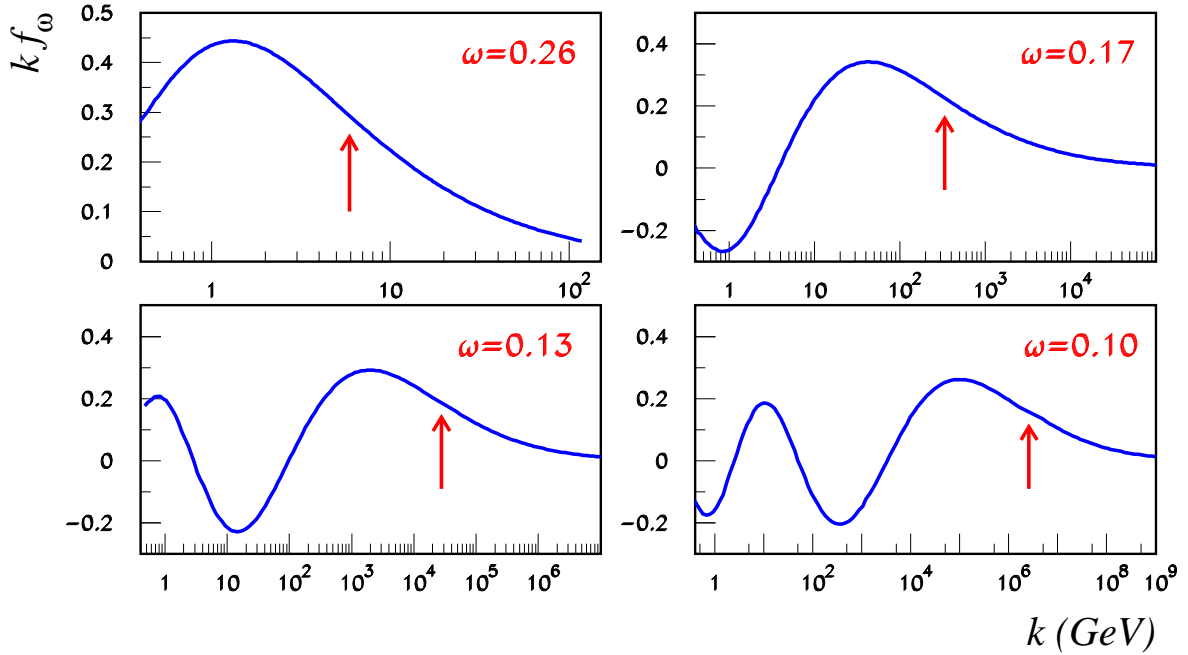,  width=18.0cm}
\vspace*{-1.3cm}
\caption{\it The first four eigenfunctions of the NLO
BFKL kernel
with running coupling and infrared phase $\eta=-0.21\pi$ at $k_0 = 0.3$~GeV.
The arrows indicate the values of $k_{crit}$.}
\label{eigenfunctions}
\end{figure}

We determined simultaneously the best-fit value of the infrared phase $\eta$ and  the coefficients,
$a_n$.  The fit was performed in the low-$x$ region, $x \le 0.01$ and for  $Q^2>4$ GeV$^2$,
so as to avoid saturation effects.  The saturation scale at HERA  was found to be
$Q^2 \sim 0.5$ GeV$^2$~\cite{KMW}, implying that saturation effects should fall below the
measurement precision for $Q^2 > 4$ GeV$^2$~\cite{aliter}.
The best fit is obtained for $\eta=-0.21\pi$, with the values for the
first four eigenvalues
and their corresponding $k_{\mathrm{crit}}$,  given in Table \ref{table:evalues} ;
\begin{table}
\begin{center}
\begin{tabular}{|c|c|c|} \hline
$n$ & $\omega$ & $k_{\mathrm{crit}}$ (GeV) \\ \hline \hline
1 & 0.26 & 5.9 \\ \hline
2 & 0.17 & 330 \\ \hline
3 & 0.13 & $2.8\times 10^4$ \\  \hline
4 & 0.10 & $2.6\times 10^6$ \\  \hline
\end{tabular}
\end{center}
\caption{\it The eigenvalues and values of $k_{crit}$ for the 4 leading eigenfunctions of
the asymptotically-free BFKL Pomeron, for $\eta=-0.21\pi$ at $k_0 = 0.3$~GeV.}
\label{table:evalues}
\end{table}
 the corresponding eigenfunctions (normalized in
the domain $k \, > \, k_0$) are shown in Fig.~\ref{eigenfunctions}.
We see that the eigenvalues indeed decrease as $n$ increases, so that
for sufficiently small $x$ the leading trajectories should be sufficient
to describe the data over any fixed range in $k$. We note also that the eigenvalues
approach each other as $n$ increases.

\begin{table}
\begin{center}
\begin{tabular}{|c|c|c|c|c|c|} \hline
Number of poles &$\chi^2/N_{df}$ & $a_1$ & $a_2$ & $a_3$& $a_4$\\ \hline \hline
1   & 3624/101       & 0.035 &    -  & -    & -    \\ \hline
2   & 264/100        & 0.029 & -0.028& -    & -    \\ \hline
3   & 91.4/99        & 0.041 & 0.055 & 0.085 & -    \\  \hline
4   & 91.3/98        & 0.042 & 0.067 & 0.11 & 0.016 \\  \hline
\end{tabular}
\end{center}
\caption{\it The qualities of fits using up to 4 poles, and the corresponding pole residues,
assuming $\eta=-0.21\pi$ at $k_0 = 0.3$~GeV.}
\label{table:evalues2}
\end{table}

Having fixed the value of $\eta$, we investigate the number of eigenfunctions required for a good description
of the HERA data for $x \le 10^{-2}$.
An overall 1-pole fit using only the leading eigenfunction has very poor quality:
$\chi^2/N_{df} = 3624/101$, though it does reproduce qualitatively the data for $x \sim 10^{-3}$,
where it is more likely to dominate over the non-leading Pomeron poles.
The quality of the overall fit improves significantly when the two first eigenfunctions are used:
$\chi^2/N_{df}=264/100$, and the 3-pole fit is excellent: $\chi^2/N_{df}=91.4/99$.
On the other hand, adding a fourth eigenfunction does not improve the fit any
further: $\chi^2/N_{df} = 91.3/98$. Since also the coefficient of the leading eigenfunction, $a_1$,
is almost the same in the 3- and 4-pole fits, in the following we consider
only the fits with 3 or less eigenfunctions.

\begin{figure}
\centering
\epsfig{file=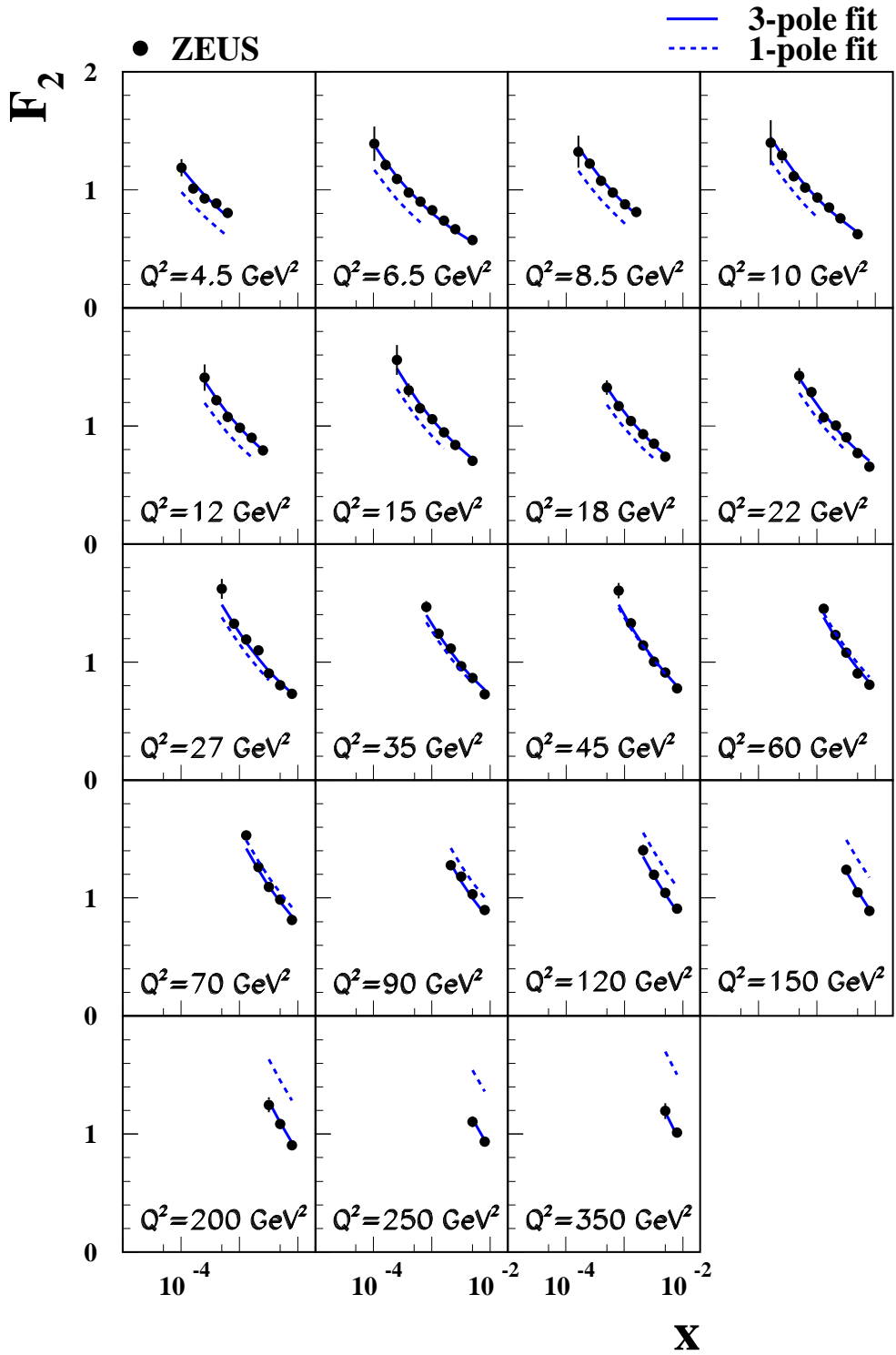, width=15cm}
\caption{\it Comparison of the HERA $F_2$ data with the 1- and 3-pole fits, shown as
dashed and solid blue lines, respectively.}
\label{polefit}
\end{figure}

\begin{figure}
\centering
\epsfig{file=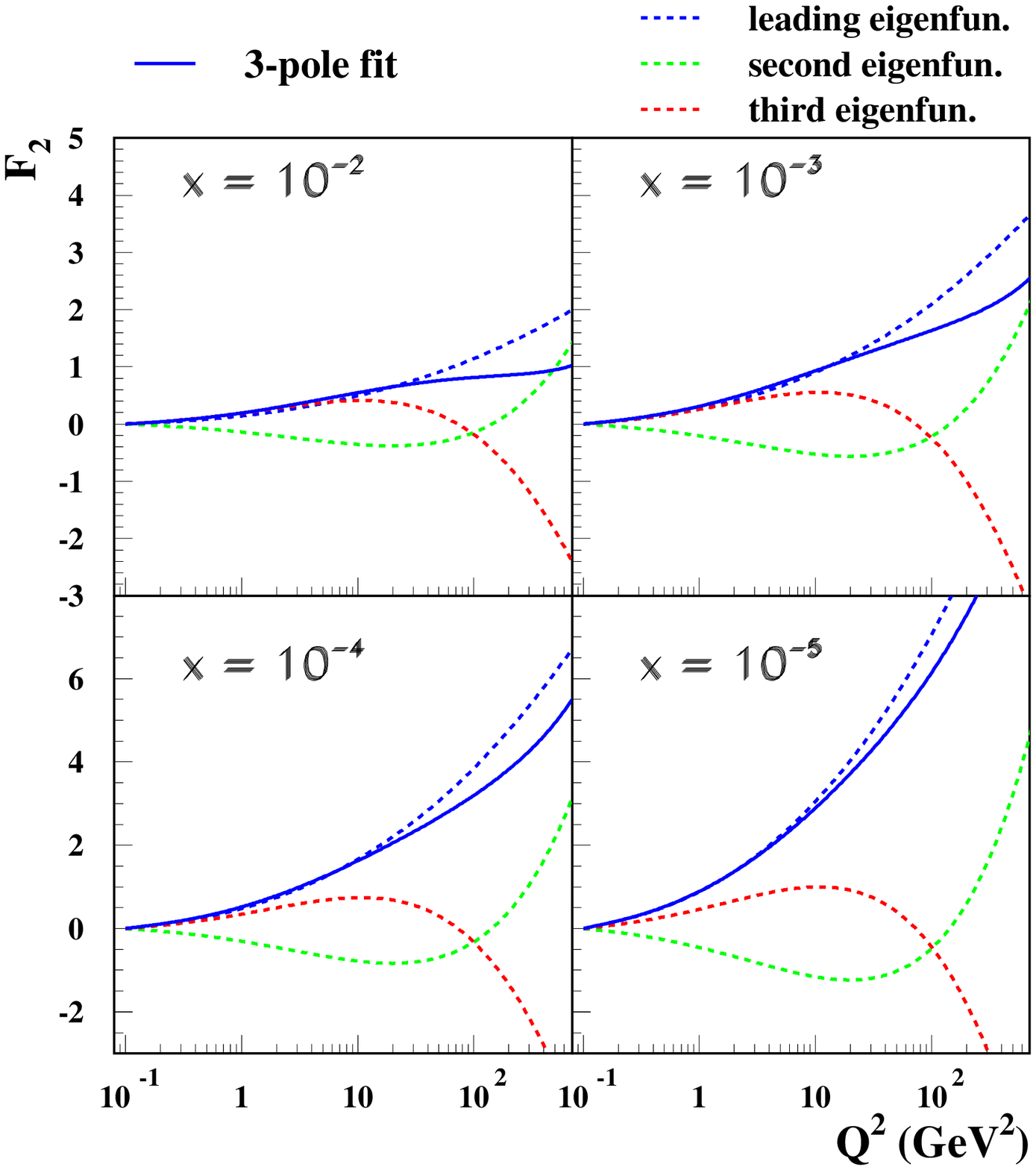, width=10cm}
\caption{\it The contributions to $F_2$ of the three eigenfunctions of the 3-pole fit.}
\label{f2comp}
\end{figure}

Fig.~\ref{polefit} compares the results
of the 1- and 3-pole fits with the measured values of $F_2$.   We see that the 3-pole fit
indeed describes the data very well, corresponding to its very good $\chi^2$. Fig.~\ref{polefit} also
displays the 1-pole fit; despite its very high  $\chi^2$, it reproduces
qualitatively the main features of the data, particularly for moderate $Q^2$. We note that
the coefficient, $a_1$, of the leading eigenfunction in the 1-pole fit is about 20\% $smaller$ than
in the 3-pole fit. This indicates that the excellent agreement of the 3-pole fit with the data is
due in part to cancellations between the different eigenfunctions. To illustrate the properties of the
3-pole fit, we show in Fig.~\ref{f2comp} the contributions to  $F_2$  from the 3 eigenfunctions
separately as
functions of the momenta $k^2$ at several characteristic $x$ values.
Fig.~\ref{f2comp} shows that, in the region of medium $Q^2$ values: $4<Q^2<20$ GeV$^2$, the contribution of the leading
eigenfunction coincides with the fitted $F_2$ curve, i.e., the
contributions of the second and third eigenfunctions cancel each other. However,
at larger $Q^2$, especially above $Q^2>100$ GeV$^2$, the fit has large cancellations between all three
components, and the leading eigenfunction cannot fit the data by itself.

\begin{figure}
\vspace*{-2.cm}
\centering
\epsfig{file=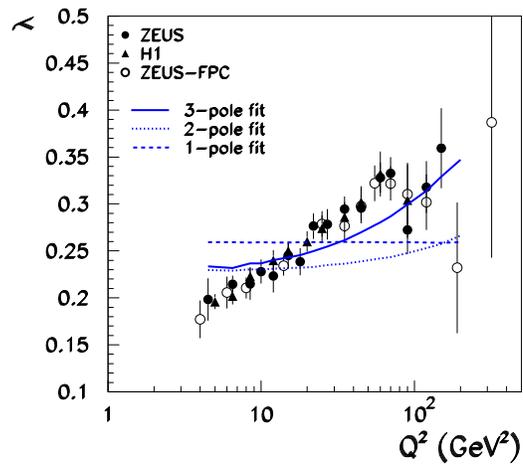, width=8cm}
\caption{\it The rate of rise  $\lambda$,  defined by $F_2 \propto (1/x)^\lambda$ at fixed $Q^2$, as determined
in the three fits and in a direct phenomenological fit to the data.}
\label{lam}
\end{figure}

Fig.~\ref{lam} compares the $Q^2$ dependence of the effective value
of the exponent $\lambda$ determined from
a phenomenological fit to the data, and as extracted from our fits
\footnote{In  Fig.~\ref{lam} we have also included recent data from Zeus \cite{Zeus-FPC}
which are fully consistent with previous data, and therefore have not been used in the fit.}.
In the case of the 1-pole fit (dashed line), $\lambda$ is identical with the leading eigenvalue:
$\omega =0.26$, and in the 2-pole fit (dotted line) the values of $\lambda$ become mostly smaller
than the leading eigenvalue. However, in the 3-pole fit, whilst the $\lambda$ values are smaller
than the leading eigenvalue for $Q^2<20$ GeV$^2$ (solid line), they become larger at higher
$Q^2$, and are closer to the values extracted form a phenomenological fit to the data.
The surprising fact that the sum of the contribution  with small eigenvalues can give a larger rate of rise
than the leading eigenvalue is due to the fact that $\lambda$ is closely connected to the logarithmic
derivative, $\lambda \approx d\log(F_2)/d\log(1/x)$. Owing to these cancellations, the logarithmic derivative
can become larger than the largest eigenvalue.
The fact that the 3-pole BFKL fit gives somewhat smaller values of $\lambda$ than the
phenomenological fit for $Q^2 \sim 20$ to 70~GeV$^2$ is closely related to the fact that the
lowest-$x$ point at each of these values of $Q^2$ lies slightly above the 3-pole fit, as seen in the
corresponding panels of Fig.~\ref{polefit}. However, the minor discrepancies for these few
points does not spoil the quality of the overall fit, which is a better measure of its validity than
the $\lambda$ plot shown in Fig.~\ref{lam}.

In summary: we obtain a very good description of the HERA low-$x$ data in a large
range of $Q^2$, from $4<Q^2<650$ GeV$^2$,
 using just  three eigenfunctions and adjust  4 free constants: the phase $\eta$ and the coefficients
$a_{1,2,3}$~\footnote{We recall that adding the fourth pole does not improve the fit.}.
Important roles are played in the fit not only by the leading eigenfunction, but also by the pattern
of cancellations between the sub-leading trajectories, which is very sensitive to the parameter
$\eta$. For this reason, the quality of the fit is also very sensitive to the value of $\eta$. Thus, for
$\eta=-0.3\pi$ the $\chi^2$  grows to 142 for the 3 pole fit (instead 91.4 at the minimum), and
at the extreme values of  $\eta=-\pi/2$ and $\eta=0$ the $\chi^2$ values are 430 and 680,
respectively. Consequently, the data determine the infrared phase quite precisely
(within the theoretical framework described above)
\footnote{ We emphasize that the value of $\eta$ is linked to the arbitrary value chosen
for the infrared scale, $k_0$, and furthermore that $\eta$ is very sensitive to the
theoretical input (e.g. the procedure employed for resumming the large NLO corrections).}:
$\eta = -0.21 \pm 0.02 \pi$
In turn, the leading eigenvalue is also precisely determined:
$\omega = 0.26\pm 0.01$.
The relatively low value of this eigenvalue is responsible for the fact that, although
the $\lambda$ plot is reproduced by the 3-pole fit only in a qualitative way, we
 obtain a very good overall fit to the data.

To our knowledge, this is the first time that the discrete asymptotically-free BFKL Pomeron
has been shown to fit the HERA data at low $x$ and high $Q^2$. As such, we believe that
it is also the first time that a parametrization of the Pomeron derived from first principles in
QCD has been confronted successfully with experimental data. A natural next step would
be to extend this comparison to include other low-$x$ HERA data, including
those on the diffractive production of vector mesons, etc.
One could also envisage the development of a BFKL
Pomeron calculus and its deployment to make predictions for both inclusive and exclusive
phenomena at the LHC.

\section*{Acknowledgements}

We express our warm appreciation for the inspiration and guidance offered to us by Lev Lipatov.
In addition, HK thanks Jochen Bartels and Leszek Motyka for useful discussions.

\end{document}